\documentclass[a4paper,fleqn]{cas-dc}
\usepackage{float}
\usepackage{hyperref}
\hypersetup{
    colorlinks=true,
    urlcolor=black,     
    citecolor=blue,     
    linkcolor=blue      
}
\usepackage{caption} 
\usepackage[numbers]{natbib}
\def\tsc#1{\csdef{#1}{\textsc{\lowercase{#1}}\xspace}}
\tsc{WGM}
\tsc{QE}
\tsc{EP}
\tsc{PMS}
\tsc{BEC}
\tsc{DE}

\ExplSyntaxOn
\cs_gset:Npn \__first_footerline: { }
\ExplSyntaxOff

\begin{document}
\let\WriteBookmarks\relax
\def\floatpagepagefraction{1}
\def\textpagefraction{.001}
\shorttitle{Wavelet Scattering Transform for Interpretable Schizophrenia}

\title [mode = title]{Wavelet Scattering Transform for Interpretable Schizophrenia Biomarker Discovery and Classification from Resting-State EEG}                      


\author[1]{Md. Taksimul Ahsan Tawhid}[orcid=0009-0002-3530-3290] 
\cormark[1] 
\fnmark[1]  
\ead{ahtawhid12345@gmail.com}


\author[1]{Nasif Ahmed Rafe}[orcid=0009-0006-7223-318X]

\cormark[1] 

\fnmark[1]  

\ead{nasifrafe229@gmail.com}


\author[2]{Alif Tahmid Priyom}[orcid=0009-0008-4491-525X] 

\ead{tahmidpriyom@gmail.com} 


\author[3]{K. M. Mustafizur Rahman} [orcid=0000-0001-8739-2470]

\ead{ mustafizbcc@eece.mist.ac.bd}

\affiliation{organization={Department of EECE, Military Institute of Science and Technology (MIST)},
                addressline={Mirpur Cantonment},
                city={Dhaka},
                postcode={1216},
                country={Bangladesh}}

\cortext[cor1]{Corresponding author(s).}
\fntext[fn1]{These authors contributed equally to this work.}
\begin{abstract}
Schizophrenia is a debilitating neuropsychiatric disorder characterized by profound cortical network dysregulation, for which objective, clinically translatable electroencephalography(EEG)-based biomarkers remain underdeveloped. Existing automated classification pipelines rely predominantly on static power spectral density features inherently blind to amplitude modulation dynamics and cross-frequency coupling, phenomena central to schizophrenia pathophysiology, while adopting epoch-level cross-validation strategies that introduce temporal data leakage, artificially inflate reported performance beyond clinically realizable levels. This study introduces a mathematically principled diagnostic framework integrating the multi-order Wavelet Scattering Transform(WST), strict Leave-One-Subject-Out (LOSO) cross-validation, and SHAP explainability for simultaneous EEG classification and biomarker discovery. Hierarchical WST coefficients capturing multi-scale amplitude modulation structure was extracted from resting-state multichannel EEG. Subject-level ANOVA with Benjamini–Hochberg false discovery rate correction identified significant biomarkers with Random Forest and SVM classifiers evaluated under strict LOSO cross-validation and subject-level majority voting. Second-order scattering coefficients encoding cross-frequency coupling dominated the discriminative biomarker set, with gamma-band features most prevalent, demonstrating that temporal amplitude modulation constitutes the primary electrophysiological signature of schizophrenia. Electrode P3 was identified as the single most discriminative site. Under rigorous subject-independent evaluation, the Random Forest achieved 90.48\% accuracy (AUC = 0.9339; sensitivity = 95.56\%). The proposed WST framework establishes a rigorous, interpretable standard for EEG-driven psychiatric biomarker discovery that can also be applicable in the detection of schizophrenia subtypes in the future.
\end{abstract}








\maketitle

\section{Introduction}

Schizophrenia is a chronic, severely debilitating neuropsychiatric disorder with a lifetime prevalence of approximately 0.7 to 1.0 percent across diverse populations and demographic groups worldwide, and around 24 million individuals are currently affected globally\cite{b1,b2}. In clinical representation, it is characterized by a heterogeneous and fluctuating triad of symptom dimensions\cite{b3}. Studies found this disorder consistently ranks among the leading causes of disability-adjusted life years in high-, middle-, and low-income countries\cite{b4},  associated with an estimated 14.5-year reduction in life expectancy compared to the general population\cite{b5}, and generates annual direct and indirect economic costs running to hundreds of billions of US dollars globally\cite{b6,b7,b8}.

Though antipsychotic pharmacotherapy can reduce positive symptoms, on average, a two-year delay occurs after the first symptoms to get proper treatment, leading to further brain damage\cite{b9,b10}. The diagnosis of schizophrenia remains mostly dependent on subjective clinical assessment as codified in\cite{b11}. Dependency solely on clinician interviews introduces vulnerabilities and may also overlap with bipolar disorder, schizoaffective disorder, drug-induced psychosis, or other medical conditions\cite{b12}. Furthermore, the current diagnostic framework is not suitable to measure any neurobiological substrate, which leads to two patients with the same diagnosis of schizophrenia having completely different brain abnormalities\cite{b13}. 

A biomarker is a characteristic that is measured as an indicator of normal biological processes, pathological processes, or biological responses to an exposure or intervention\cite{b14,b15}. In the context of schizophrenia, biomarkers are often studied using brain structure and activities in mircro levels\cite{b16}, neurotransmitters \& receptors, genetics \& RNA, and growth factors\cite{b17}. Recent studies found the effectiveness of using EEG data as biomarkers for an experimental glutamatergic agent in patients with schizophrenia\cite{b18}. Schizophrenia is increasingly understood as a disorder of neural temporal dynamics\cite{b19}, which makes EEG biomarker more effective because of its high temporal resolution.

In the past two decades, the classification of this abnormality from EEG signals through machine learning has become one of the best techniques. Convolutional neural network (CNN) architectures have been prominently applied in\cite{b20}, achieving high classification accuracy. With batch normalization and dropout regularization reported further improvements in schizophrenia EEG datasets\cite{b21}. Attention mechanisms and transformer-based architectures have recently been explored for long-range temporal dependency modelling in EEG\cite{b22}. Graph neural networks (GNNs) operating on electrode-level functional connectivity graphs\cite{b23} have provided a principled framework for jointly exploiting spectral features. In the domain of time-frequency feature, Khare et al.\cite{b24} employed tunable-Q wavelet transform decomposition with entropy-based feature extraction, while Baygin et al.\cite{b25} applied Collatz pattern-based features from wavelet sub-bands, both reporting strong cross-validated performance and highlighting the discriminative advantage of multi-scale time-frequency representations over purely spectral methods. Hybrid frameworks combining handcrafted feature extraction with deep classifiers have additionally been proposed to balance interpretability and representational capacity\cite{b26}. A consistent methodological concern traverses as performance is evaluated at the epoch level using k-fold cross-validation without subject-level partitioning, inflating temporal data leakage\cite{b27}; feature representations remain largely anchored to first-order spectral energy, precluding access to amplitude modulation and cross-frequency coupling structure.

The Wavelet Scattering Transform (WST), introduced by Mallat\cite{b28}and rigorously formalized by Bruna and Mallat\cite{b29}, offers a uniquely compelling framework for EEG feature extraction that simultaneously addresses the representational, stability, and interpretability requirements. The WST has been successfully applied to recognition and classification in biomedical signals\cite{b30,b31}. Complementarily, the adoption of SHapley Additive exPlanations(SHAP) provides mathematically rigorous, additive feature attributions for any model, enabling per-subject interpretability of classifier decisions that are directly mappable to individual scattering coefficients, electrode positions, and frequency bands\cite{b32,b33,b34}.

To address the absence of subject-independent evaluation, cross-validation schemes, and the inability of conventional spectral features to capture amplitude modulation and cross-frequency coupling structure, the present study proposes an integrated EEG-based schizophrenia classification and biomarker identification framework through Wavelet Scattering Transform. By exploiting the multi-order temporal representations of the WST, the framework simultaneously captures frequency-band energy and amplitude modulation structure within a single mathematically principled feature space, evaluated under Leave-One-Subject-Out cross-validation to ensure genuine subject-independent generalization. Statistical biomarker identification is conducted at the subject level to prevent pseudo-replication, and SHAP-based explainability analysis is incorporated not only to provide per-subject interpretability of classifier decisions but to independently validate the identified biomarkers through cross-methodological consistency, a form of joint evidence that has been largely absent from the prior literature.

\section{Methodology}\label{sec2}
\subsection{Dataset}
Publicly available dataset from kaggle has been employed in this study on adolescents consist of 84 subjects, where 39 of the subjects were healthy controls, and 45 were affected with schizophrenia. Recordings were acquired using a 16-channel electrode following the international 10–20 system, with electrode placements at F7, F3, F4, F8, T3, C3, Cz, C4, T4, T5, P3, Pz, P4, T6, O1, and O2 \cite{b35,b36}. Each subject was sampled at a sampling rate of 128 Hz, and the electrical activities were represented in µV. 

\begin{figure*}[t]
    \centering
    \includegraphics[width=0.9\textwidth, height=7.5cm, keepaspectratio=true]{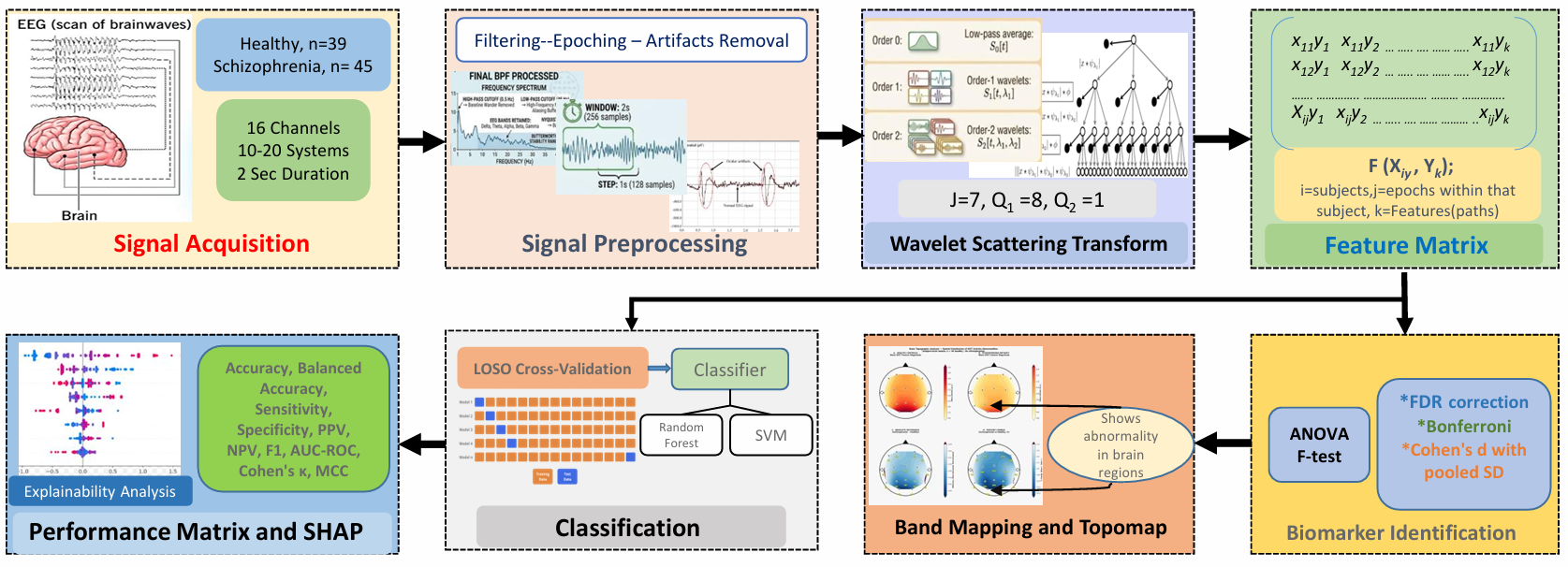}
    \caption{Overview of the proposed methodological workflow for EEG-based schizophrenia}
    \label{fig:Figure__1}
\end{figure*}

\subsection{Signal Preprocessing} 
A two-stage filtering pipeline was applied independently to each of the 16 channels across all the subjects. At first, a second-order infinite impulse response (IIR) notch filter was applied at 50 Hz to suppress power line interfaces, which constitute a primary source of non-neural signals in clinical EEG recordings. In the following step of filtering, a fourth-order zero-phase Butterworth bandpass filter was applied to retain EEG frequencies within the physiologically relevant range of 0.5–45 Hz\cite{b37}. The cutoff of 0.5 Hz effectively removes slow DC drift and electrode polarization artifacts, while the upper cutoff of 45 Hz was deliberately set below the Nyquist frequency of 64 Hz to ensure numerical stability of the Butterworth filter\cite{b38}. Then the preprocessed signal is segmented into fixed-length, temporally overlapping epochs using a sliding window approach\cite{b39,b40}. The e-th epoch of subject i at channel c is formally defined as:

\begin{equation}
    \tilde{x}_{i,c}^{(e)}(\tau)
    = \tilde{x}_{i,c}(t_{e} + \tau);
    \quad
    \tau \in \{0, 1, \ldots, L-1\}
    \label{eq:epoch_def}
\end{equation}
 
\begin{equation}
    \hfill t_{e} = (e - 1) \times\frac{L}{2} \hfill
    \label{eq:epoch_start}
\end{equation}

where i, c, and L are the total number of subjects, channels and samples respectively.

The preprocessed signals after filtering showed per-channel standard deviations of 150–450 µV. So, an adaptive z-score threshold was used, which normalizes each sample relative to the subject- and channel specific signal distribution, thereby rendering the rejection criterion invariant to absolute amplitude scale\cite{b41}. A methodological workflow  of the proposed EEG-based schizophrenia detection is presented in Fig.~\ref{fig:Figure__1}.

\subsection{Wavelet Scattering Transform (WST)}
\begin{figure*}[t]
    \centering
    \includegraphics[width=0.9\textwidth, height=7.5cm, keepaspectratio=true]{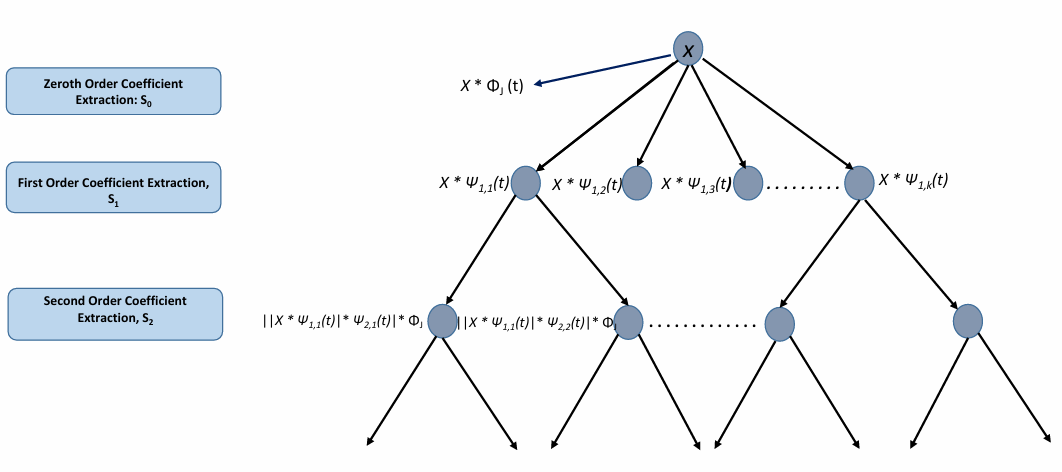}
    \caption{Hierarchical representation of multi-order WST coefficient extraction across different wavelet frequencies.}
    \label{fig:Figure__2}
\end{figure*}

Wavelet scattering Transform is done in the preprocessed epoched signal to extract three levels of coefficients, a hierarchical representation is shown in  Fig. ~\ref{fig:Figure__2}.. These three levels of coefficients carry the following meaning:

\textbf{Zeroth-order ($S_{0}$):} 
\begin{equation}
    \hfill S_0 x(t) = x \star \phi_J(t) \hfill
\end{equation}
Here, x is the EEG epoch signals; $\phi_J(t)$ real-valued lowpass averaging filter at invariance scale $J$, with temporal support $2^J$ samples. In neural signals, $S_0$ is the local DC baseline or Slow Cortical Potential (SCP). It represents the ultra-slow, averaged electrical drift of large populations of neurons over the time window defined by $2^J$.

\textbf{First-order ($S_{1}$):} 

\begin{equation}
    \hfill S_1 x(t, \lambda_1) = |x \star \psi_{\lambda_1}| \star \phi_J(t)\hfill
    \label{eq:wst_u1}
\end{equation}

$\psi_{\lambda_1}(t)$complex-valued analytic Morlet wavelet at wavelet path $\lambda_1$.  It acts like a robust, deformation-stable spectrogram. It represents how much energy is present at frequency $\lambda_1$ over time.

\textbf{Second-order ($S_{2}$):} 

\begin{equation}
\hfill
    S_2 x(t, \lambda_1, \lambda_2) = \lvert \lvert x \star \psi_{\lambda_1} \rvert \star \psi_{\lambda_2} \rvert \star \phi_J(t)
\hfill
\end{equation}

This captures how brain rhythms interact with each other and how fast they change. It measures the amplitude modulation occurring inside a neural frequency band. If the first wavelet $\psi_{\lambda_1}$ isolates a high-frequency band and the second wavelet $\psi_{\lambda_2}$ analyzes it at a lower frequency, $S_2$ physically quantifies how the fast rhythm is being driven, pulsed, or modulated by the slow rhythm.

\textbf{The Scale Parameter(J):} This sets the temporal window ($2^J$) over which the brain activity is averaged. For  2-second EEG epochs, the invariance scale parameter was optimized at $J = 7$, establishing a 1.0-second temporal window that satisfies the boundary constraint \cite{b28}. This specific scale is the optimal choice because it maximizes deformation stability against pathological phase jitters\cite{b42} without completely sacrificing temporal dynamics.$J=7$ yields each epoch in two time positions per path. Thus, it balances between robust feature invariance and the capacity to track macro-structural state transitions critical for schizophrenia classification\cite{b43}.

\textbf{The Quality Factor(Q):} The configuration Q = (8, 1) was used, as $Q_1$= 8, to provide high frequency resolution for isolating closely spaced pathological neural oscillations. Again, $Q_2$ = 1 maximizes temporal localization in the second layer to accurately trap sudden, non-stationary micro-transients characteristic of schizophrenia dynamics\cite{b44}.

\subsection{Biomarker Identification}
The features from WST is then used to make a feature vector. The complete epoch-level feature vector concatenating all channels:

\begin{equation}
\hfill f_i^{(e)} = \left[ Sx_{i,1}^{(e)} ; Sx_{i,2}^{(e)} ; \dots ; Sx_{i,16}^{(e)} \right] \hfill
\end{equation} 
where $f_i^{(e)}$ is the complete feature vector for epoch $e$ of subject $i$, and $Sx_{i,c}^{(e)}$ is the scattering representation of epoch $e$, subject $i$, channel $c$.

The epoch-level feature matrix is constructed by stacking feature
vectors of all retained epochs. For all statistical analyses, a subject level feature matrix is derived by computing the mean feature vector across each subject.

All statistical hypothesis testing is conducted exclusively to prevent pseudo replication, the error of treating non-independent, temporally overlapping epochs from the same subject as statistically independent observations, which artificially inflates degrees of freedom and yields
unreliable p-values\cite{b45}.

\subsubsection{Subject-Level ANOVA F-Test}
A one-way ANOVA F-test was performed independently for each of the WST features on the subject-level matrix\cite{b46}. For each feature $d \in \{1, 2, \dots, D=\text{Number of features}\}$, the cohort-specific mean vectors of healthy control and schizophrenia groups were computed alongside the grand mean of the total population. Feature variance was partitioned by deriving the between-group sum of squares ($SS_B^{(d)}$) and within-group sum of squares ($SS_W^{(d)}$), which quantify class divergence and intra-group dispersion, respectively. The statistical significance of each feature was evaluated using the $F$-statistic, formulated with degrees of freedom $df_{\text{between}} = 1$ and $df_{\text{within}} = \text{Total Subjects} - 2 = 82$:

\begin{equation}
\hfill  F^{(d)} = \frac{SS_B^{(d)} / 1}{SS_W^{(d)} / 82} \hfill
\end{equation}

\subsubsection{Multiple Testing Correction}

Multiple testing correction is essential to control the false discovery rate. The Benjamini-Hochberg (BH) procedure\cite{b47} was selected as the primary correction method. The $p$-values corresponding to $F^{(d)}$ be sorted in ascending order as  $p_{(1)} \le p_{(2)} \le \dots \le p_{(D)}$. The BH rejection rule at target false discovery rate $q = 0.05$ is:

\begin{equation}\hfill
\text{Reject }   H_0^{(d)}; \quad \text{if} \quad p_{(d)} \le \left( \frac{d}{D} \right) \times q \hfill
\end{equation}

The significant feature index set at $q < 0.05$:

\begin{equation}\hfill
D_{\text{sig}} = \left\{ d : p_{(d)} \le \frac{d \times 0.05}{D} \right\} \hfill
\end{equation}

Bonferroni-corrected significance threshold; more conservative than BH-FDR; features surviving this threshold are reported as supplementary strongly significant biomarkers\cite{b48}:

\begin{equation}
\hfill
\alpha_B = \frac{0.05}{D} = 8.88 \times 10^{-6}
\hfill
\end{equation}

\textbf{Effect Size Estimation:}
To quantify the magnitude of cross-cohort divergence for each selected feature $d \in D_{\text{sig}}$, the standardized effect size was evaluated using Cohen’s $d$ , utilizing a weighted pooled standard deviation to rigorously account for the unequal group sizes\cite{b49}.

\subsection{Topographic Brain Mapping}
Each of the paths was assigned to a standard EEG frequency band based on its centre frequency. Spatial topographic maps were generated by projecting per-electrode statistics onto a two-dimensional scalp model via standard 10-20 coordinates, applying cubic spatial interpolation over a $400 \times 400$ grid\cite{b50}. Hemispheric asymmetry was quantified using the Lateralization Index ($LI$):
\begin{equation}
\hfill
LI = \frac{\bar{\mu}_L - \bar{\mu}_R}{|\bar{\mu}_L| + |\bar{\mu}_R|}
\hfill
\end{equation}

where $\bar{\mu}_L$ and $\bar{\mu}_R$ are the mean absolute Cohen's $d$ values averaged over the seven left-hemisphere electrodes $\{\text{F7}, \text{F3}, \text{T3}, \text{C3}, \text{T5}, \text{P3}, \text{O1}\}$ and seven right-hemisphere electrodes $\{\text{F8}, \text{F4}, \text{T4}, \text{C4}, \text{T6}, \text{P4}, \text{O2}\}$, respectively. $\text{LI} \in [-1, +1]$, with $\text{LI} < 0$ indicating left-dominant disruption and $\text{LI} > 0$ indicating right-dominant disruption. Statistical significance of hemispheric and anterior-posterior asymmetries was assessed using the non-parametric Mann-Whitney $U$ test\cite{b51}, which makes no assumption of normality and is appropriate for the small electrode-group sample sizes involved.

\subsection{Classification}

Achieving clinically relevant classification requires a cross-validation strategy that strictly enforces subject level separation between training and test data. Leave-One-Subject-Out (LOSO) cross-validation was therefore adopted as the evaluation framework\cite{b52,b53}, wherein each of the 84 subjects serves as the test set exactly once while the remaining 83 subjects constitute the training set, yielding 84 non-overlapping folds. This is the standard for EEG classification studies\cite{b54}, as it guarantees no temporal data leakage that systematically inflates performance in standard k-fold cross-validation\cite{b27,b55}. Within each fold, a Standard Scalar was fitted exclusively on training epochs and applied to both training and test sets, preventing distributional information from the test subject from influencing model fitting. Two classifiers were evaluated. A Random Forest (RF) with 200 trees, maximum depth 20, and balanced class weights was trained on all the dimensional WST feature space. The RF was selected as the primary classifier because it naturally handles high-dimensional feature spaces, provides inherent feature importance estimates that can be cross-referenced with the statistical biomarker analysis, and is robust to class imbalance through the balanced weighting scheme\cite{b56}. An SVM with RBF kernel was trained on the FDR-significant features identified. Since classification operates at the epoch level but clinical utility requires a single diagnosis per patient, subject-level predictions were obtained through majority voting across all retained test epochs. The mean predicted probability across epochs was used as a continuous subject-level score for ROC-AUC computation. This majority vote strategy has been shown to improve subject-level decision stability over single-epoch prediction in EEG classification studies\cite{b57,b58}.

\subsection{SHAP Explainability Analysis}

High classification accuracy alone is insufficient for a biomarker discovery framework. SHAP (SHapley
Additive exPlanations) analysis \cite{b33,b34} was applied to
provide per-subject feature attribution and to cross-validate the
statistical biomarkers through an entirely independent data-driven approach.
SHAP values were computed using the Tree Explainer algorithm
\cite{b33} on the subject-level feature matrix.
Tree Explainer exploits the tree structure of the RF ensemble to
compute exact Shapley values in polynomial time without Monte Carlo
approximation or background sampling, making it both computationally
efficient and mathematically exact for tree-based models.

\section{Results}\label{sec3}
\begin{figure*}[t!]
    \centering
    \parbox{0.49\textwidth}{
        \centering
        \includegraphics[width=0.5\textwidth]{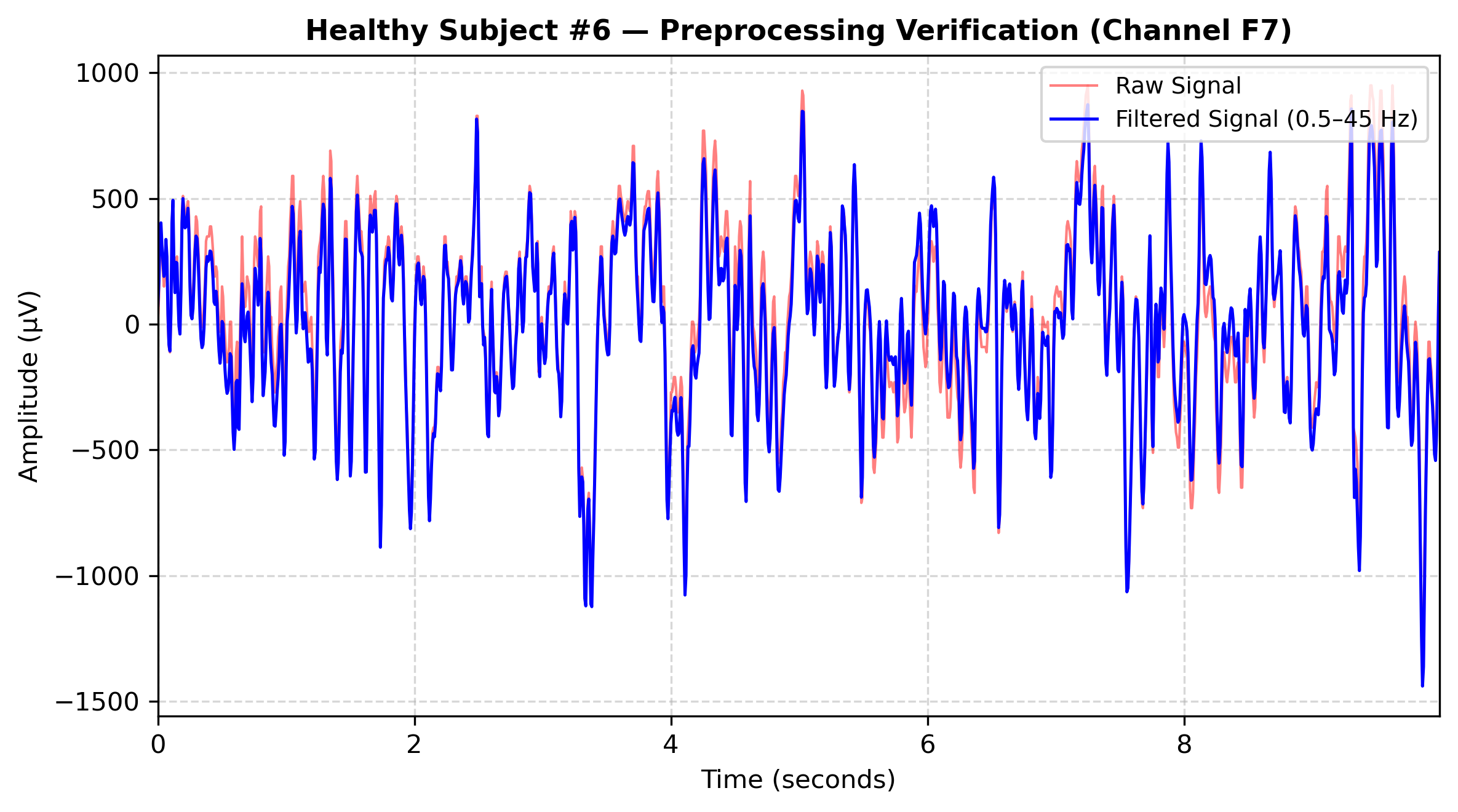}
        \centerline{\footnotesize (a) Raw vs. filtered EEG signal comparison.}
        \label{fig:3_a}
    }
    \hfill
    \parbox{0.49\textwidth}{
        \centering
        \includegraphics[width=0.5\textwidth]{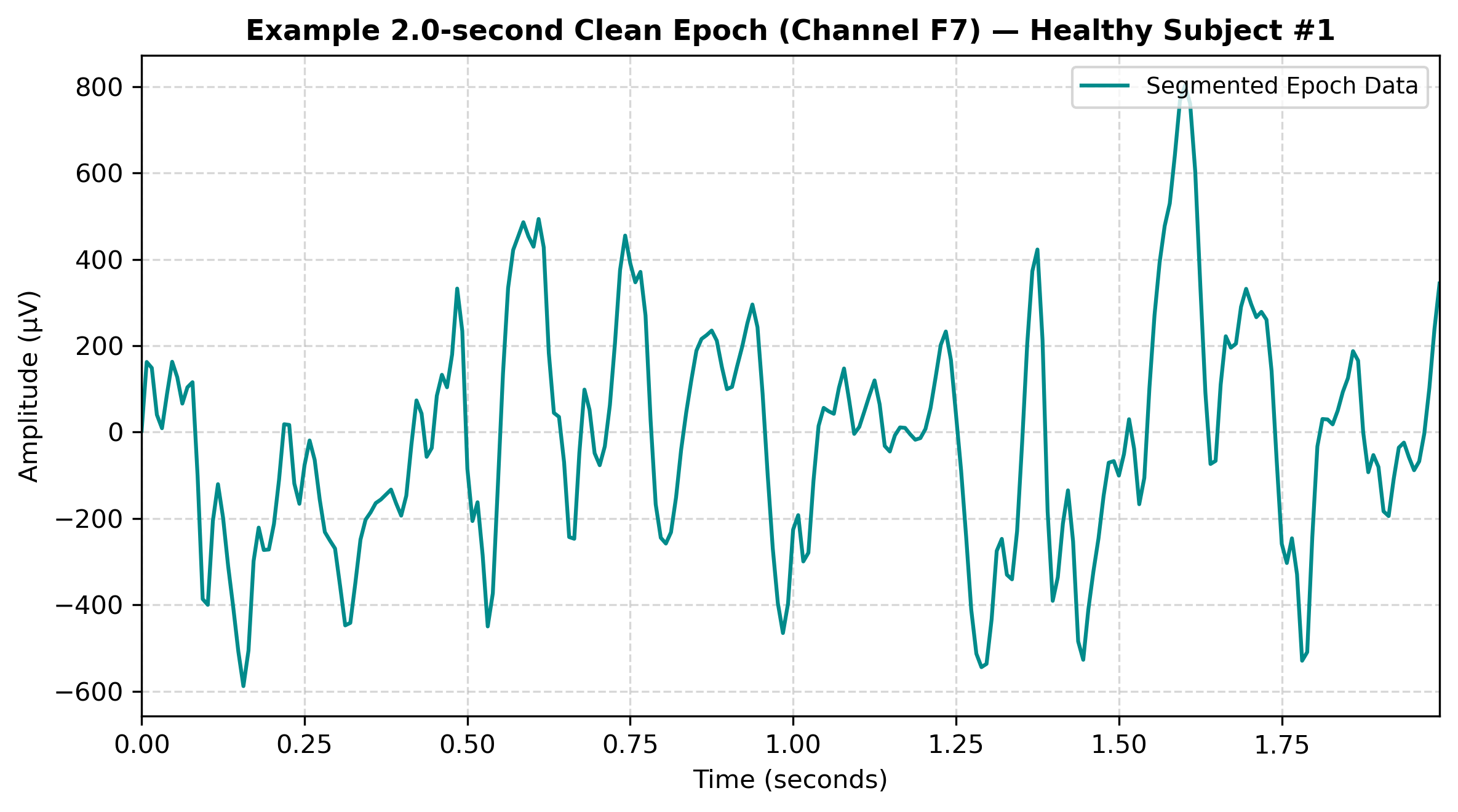}
        \centerline{\footnotesize (b) First 2-second segmented clean epoch.}
        \label{fig:epoch_f7}
    }
    
    \vspace{0.3cm}
    \caption{EEG signal preprocessing and epoch segmentation stages for channel F7: (a) raw versus zero-phase Butterworth bandpass filtered profiles, and (b) isolated 256-sample stationary epoch tracking localized voltage fluctuations.}
    \label{fig:Figure__3}
\end{figure*}

Filtering effectively mitigated noise artifacts from the raw EEG data. This continuous signal was subsequently segmented into 2-second epochs with a 50\% overlap, the raw vs filtered version of F7 channel and 2 seconds epoched signal are shown in Fig.~\ref{fig:Figure__3}.

Following z-score artifact rejection with threshold, a
total of 4,650 epochs were retained across all 84 subjects.

Using kymatio\cite{b59}, the Wavelet Scattering Transform was implemented using a Morlet wavelet as the mother wavelet. With optimization parameters set to J = 7, Q = (8, 1),
and temporal averaging generated 176 scattering paths per epoch,
comprising 1 zeroth-order ($S_0$), 46 first-order($S_1$), and 129
second-order ($S_2$) paths. With 2 time-averaged output values per
path, thus each epoch contributed 352 scattering coefficients.

\subsection{Biomarker Analysis}
Subject-level ANOVA F-testing across all WST features and the correction methods were applied, yielding the results depicted in Table~\ref{tab:method_comparison}.

\begin{table*}[t!]
\centering
\caption{Comparison of different multiple testing correction methods.}
\label{tab:method_comparison}
\begin{tabular}{llcc}
\toprule
Method & Threshold & Significant & \% of Total \\
\midrule
Uncorrected & $p < 0.050$ & 1,845 & 32.8\% \\
Bonferroni (FWER) & $p < 8.88 \times 10^{-6}$ & 27 & 0.5\% \\
FDR (Benjamini--Hochberg) & $q < 0.050$ & 1,255 & 22.3\% \\
FDR (Benjamini--Yekutieli) & $q < 0.050$ & 438 & 7.8\% \\
\bottomrule
\end{tabular}
\end{table*}

A highly conservative subset of 27 features simultaneously survived strict Bonferroni correction. Spatial mapping of the top 20 biomarkers revealed a highly localized topological pattern: 12 features mapped to electrode P3 (left parietal), 5 to F4 (right frontal), and the remainder to P4 and Pz (midline parietal). Notably, these top-tier biomarkers comprised both first- and second-order scattering coefficients, demonstrating that both localized spectral energy and cross-frequency amplitude modulation dynamics contribute fundamentally to cohort divergence.

\begin{figure*}[t!]
    \centering
    \parbox{0.49\textwidth}{
        \centering
        \includegraphics[width=0.5\textwidth]{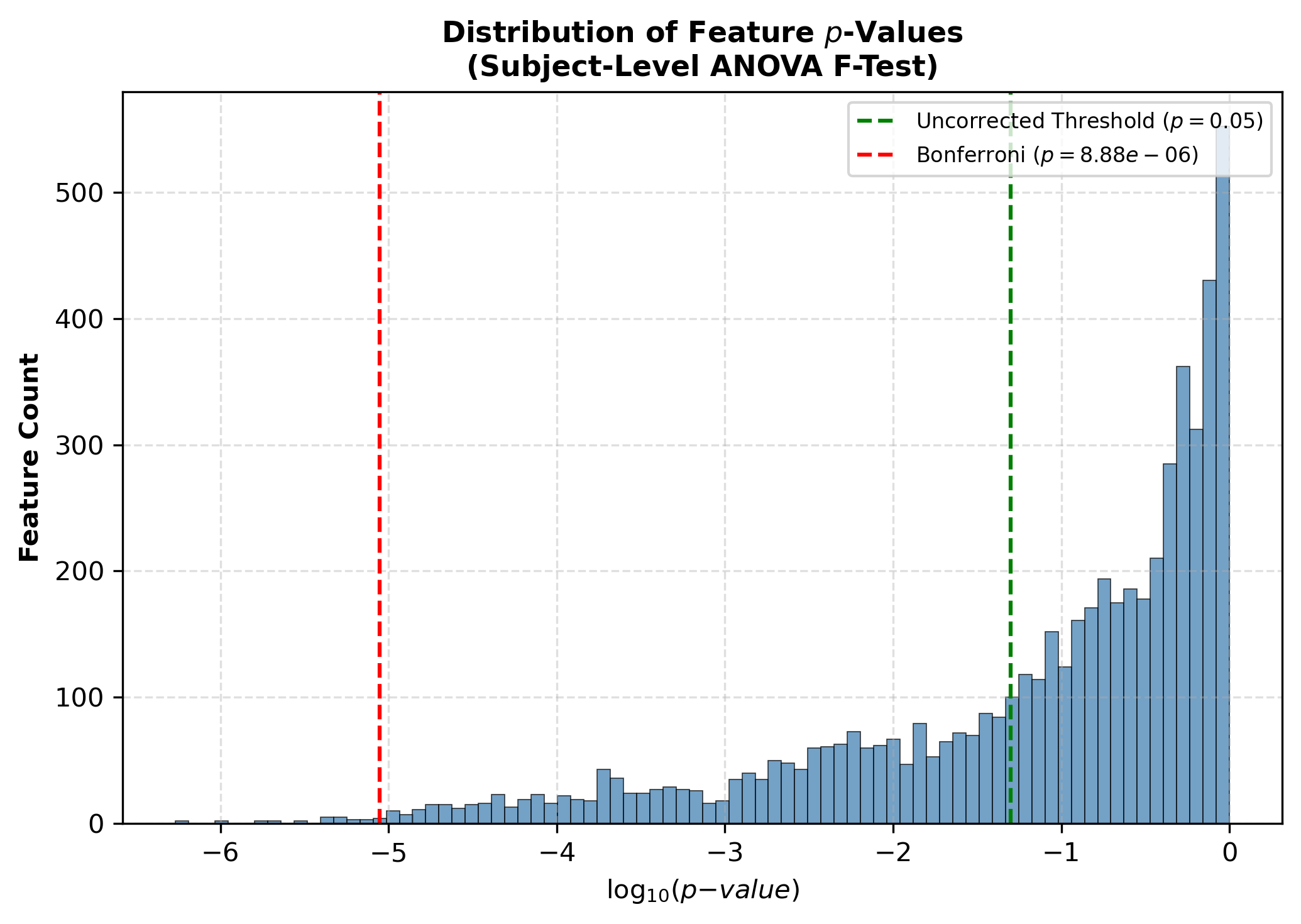}
        \centerline{\footnotesize (a) Feature $p$-value distribution.}
        \label{fig:step1_p_dist}
    }
    \hfill
    \parbox{0.49\textwidth}{
        \centering
        \includegraphics[width=0.5\textwidth]{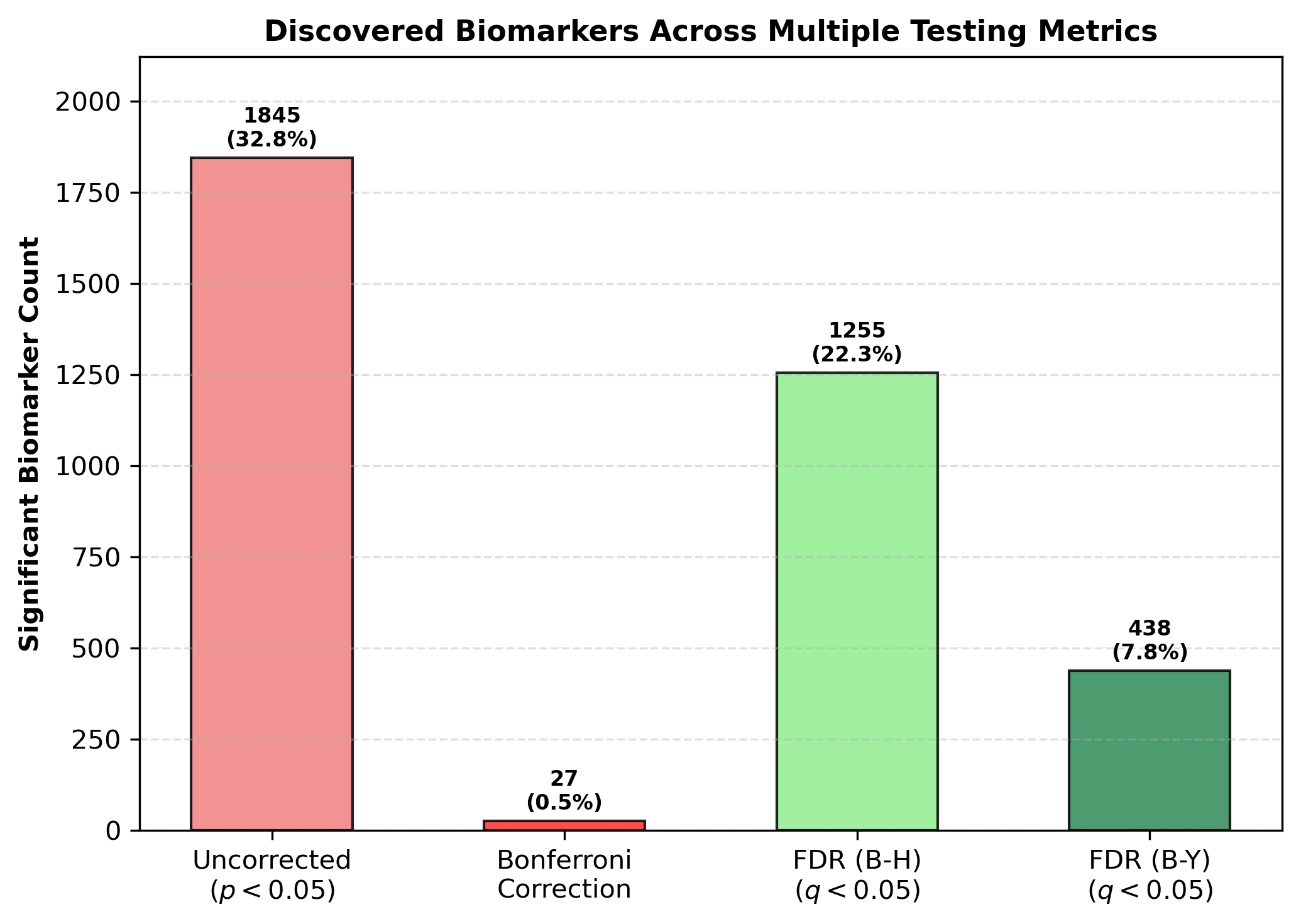}
        \centerline{\footnotesize (b) Statistical correction yield comparison.}
        \label{fig:step1_yield_comp}
    }
    
    \vspace{0.4cm} 
    
    \parbox{0.49\textwidth}{
        \centering
        \includegraphics[width=0.5\textwidth]{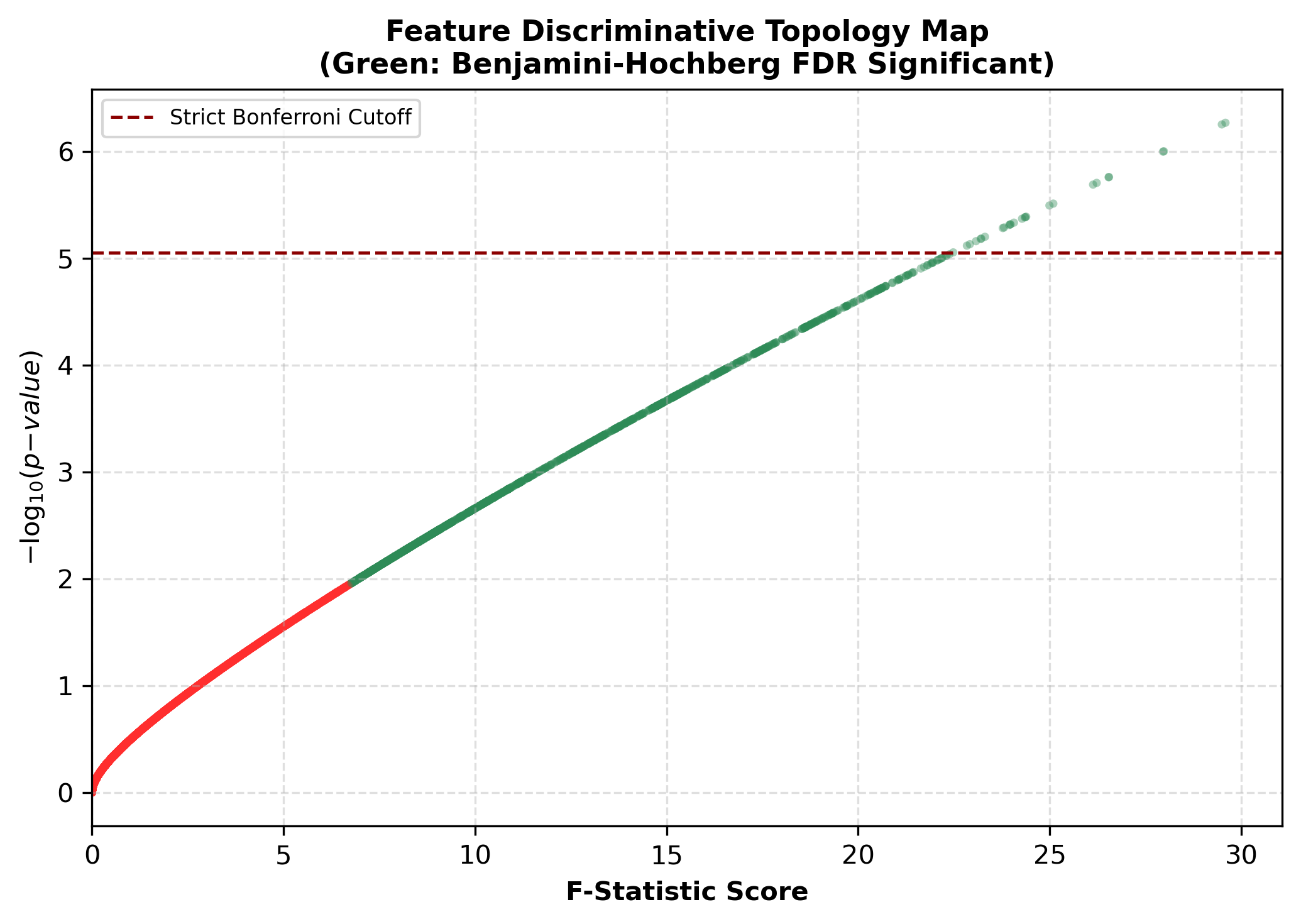}
        \centerline{\footnotesize (c) Feature discriminative topology map.}
        \label{fig:step1_volcano}
    }
    \hfill
    \parbox{0.49\textwidth}{
        \centering
        \includegraphics[width=0.5\textwidth]{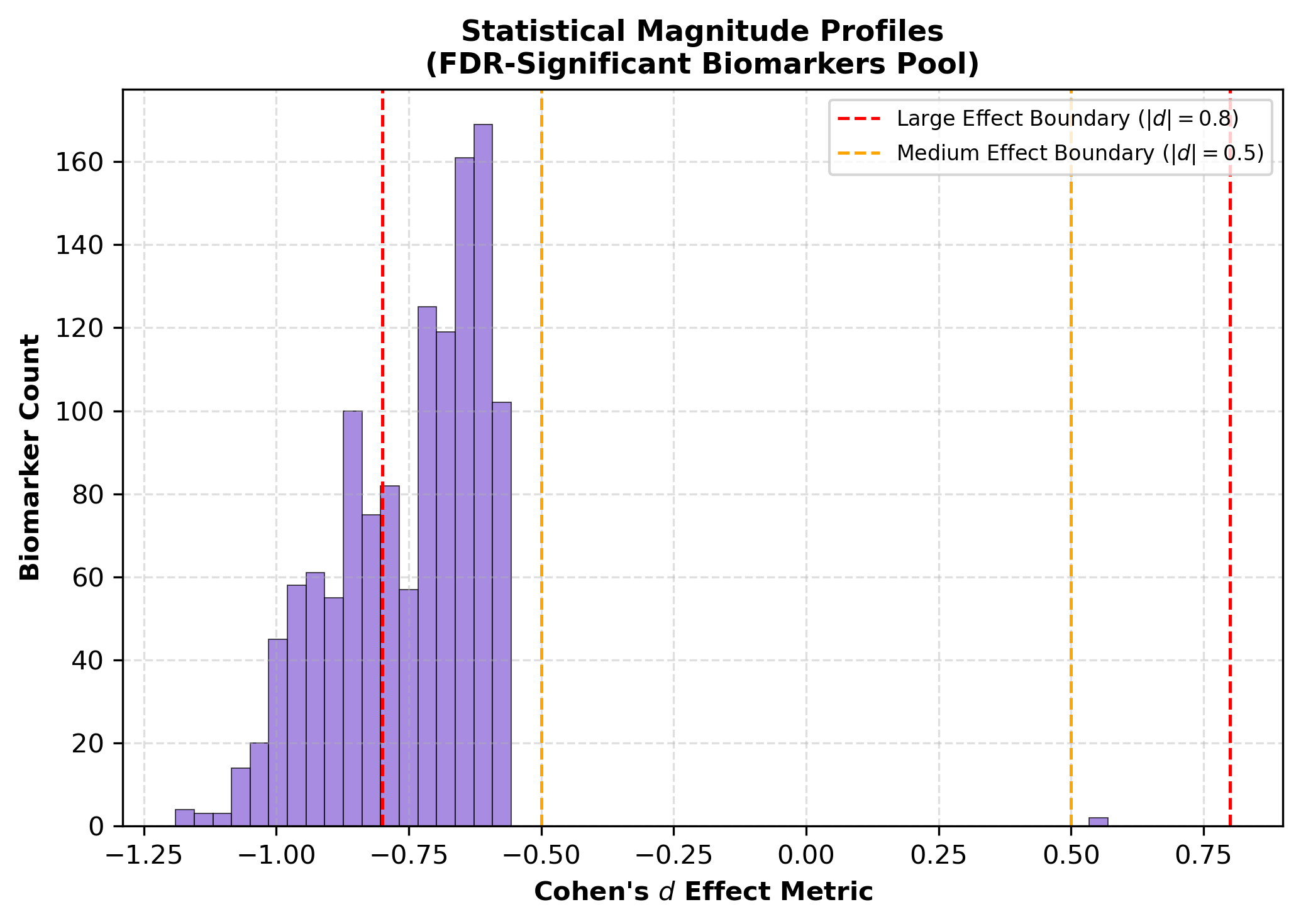}
        \centerline{\footnotesize (d) Statistical magnitude profiles (Cohen's $d$).}
        \label{fig:step1_effect_size}
    }
    
    \vspace{0.3cm}
    \caption{Statistical biomarker identification and multiple testing correction profiles for subject-level feature analysis: (a) distribution of raw feature $p$-values against uncorrected and Bonferroni thresholds, (b) total discovery yield comparison demonstrating strictness across standard multiple testing adjustments, (c) volcano topology mapping feature discriminative scores against $-\log_{10}(p\mathit{-value})$ highlights where Benjamini-Hochberg FDR adjustments maintain statistical power, and (d) Cohen's $d$ effect-size density estimation profiling the clinical magnitude of discovered schizophrenia biomarkers.}
    \label{fig:clinical_biomarkers_step1}
\end{figure*}

Effect size analysis of the 1,255 FDR-significant features revealed exclusively large ($|d| > 0.8$; $n = 443$, 35.3\%) and medium ($0.5 < |d| \le 0.8$; $n = 812$, 64.7\%) effect sizes; no small effects survived correction. Crucially, 99.8\% (1,253) of these biomarkers exhibited negative Cohen’s $d$ values, indicating a pervasive, systematic reduction in wavelet scattering energy within the schizophrenia cohort. The top-ranked biomarker (P3) demonstrated a profound effect ($d = -1.190$), translating to a 28.7\% drop in mean scattering activity. A comparison of statistical method is depicted in Fig.~\ref{fig:clinical_biomarkers_step1}.

\subsection{Spectral and Structural Feature Distributions}
\begin{figure*}[t!]
    \centering
    \parbox{0.32\textwidth}{
        \centering
        \includegraphics[width=0.32\textwidth]{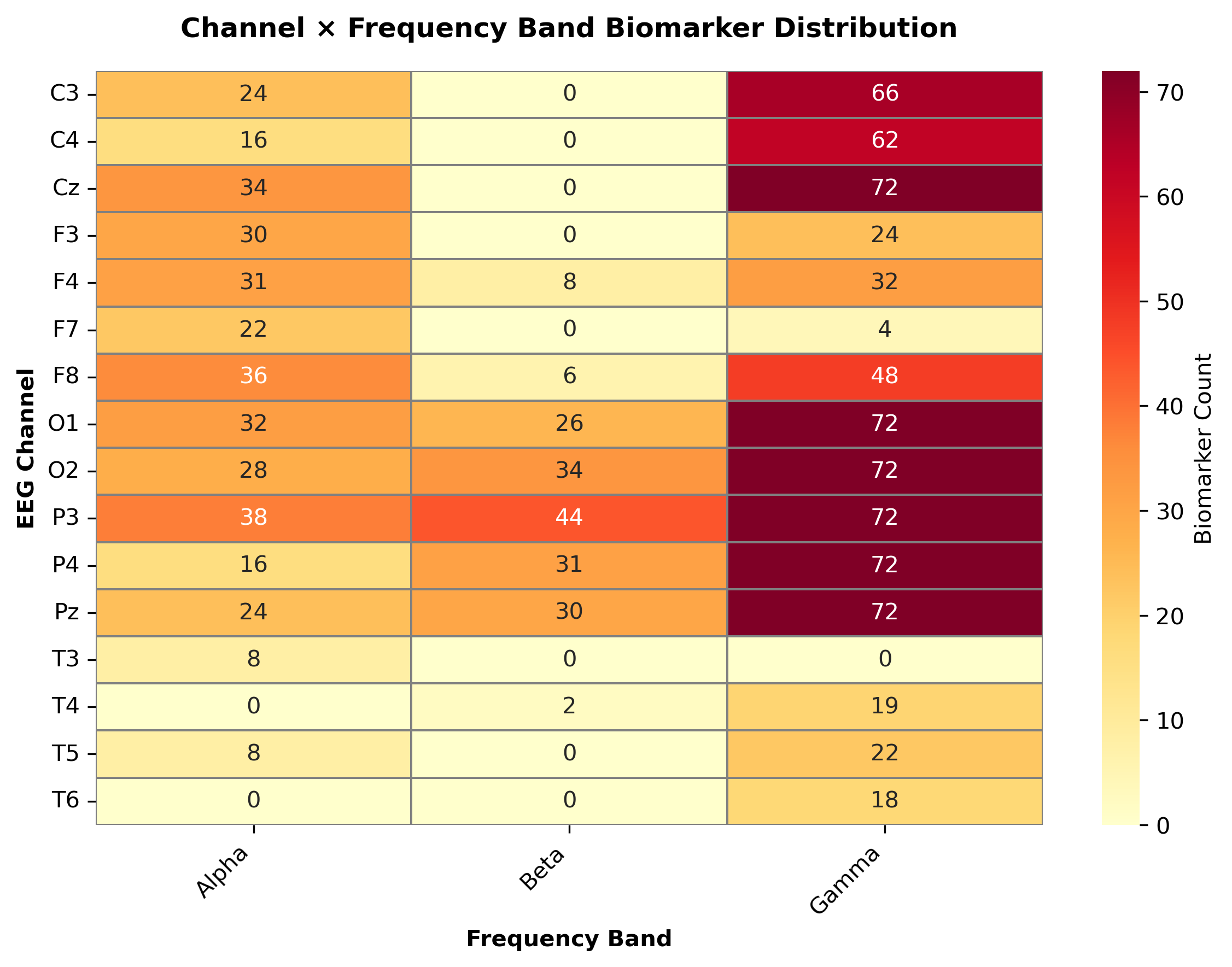}
        \centerline{\footnotesize (a) Spatial channel discovery density.}
        \label{fig:5_a}
    }
    \hfill
    \parbox{0.32\textwidth}{
        \centering
        \includegraphics[width=0.35\textwidth]{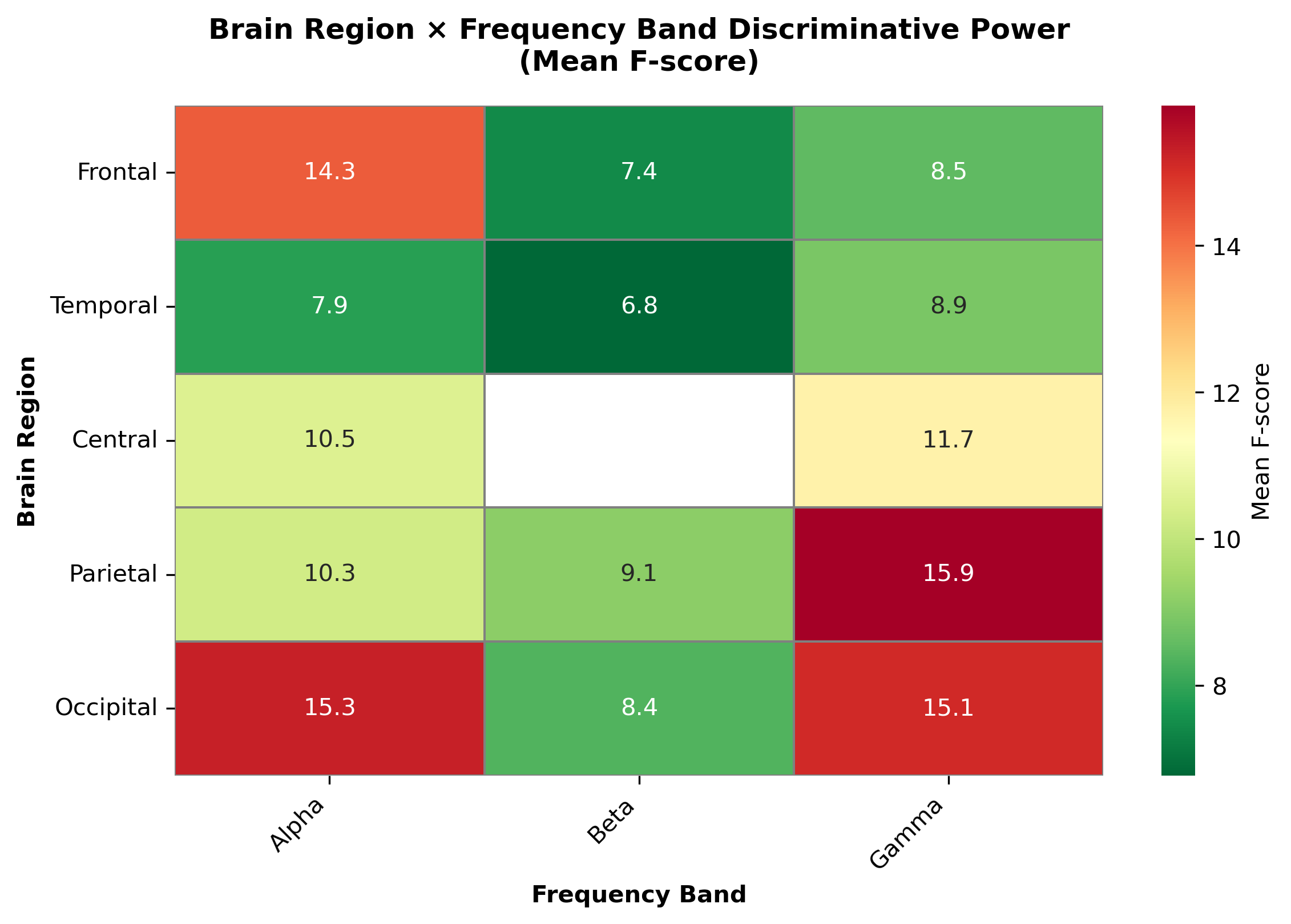}
        \centerline{\footnotesize (b) Regional mean $F$-score weight profile.}
        \label{fig:5_b}
    }
    \hfill
    \parbox{0.32\textwidth}{
        \centering
        \includegraphics[width=0.31\textwidth]{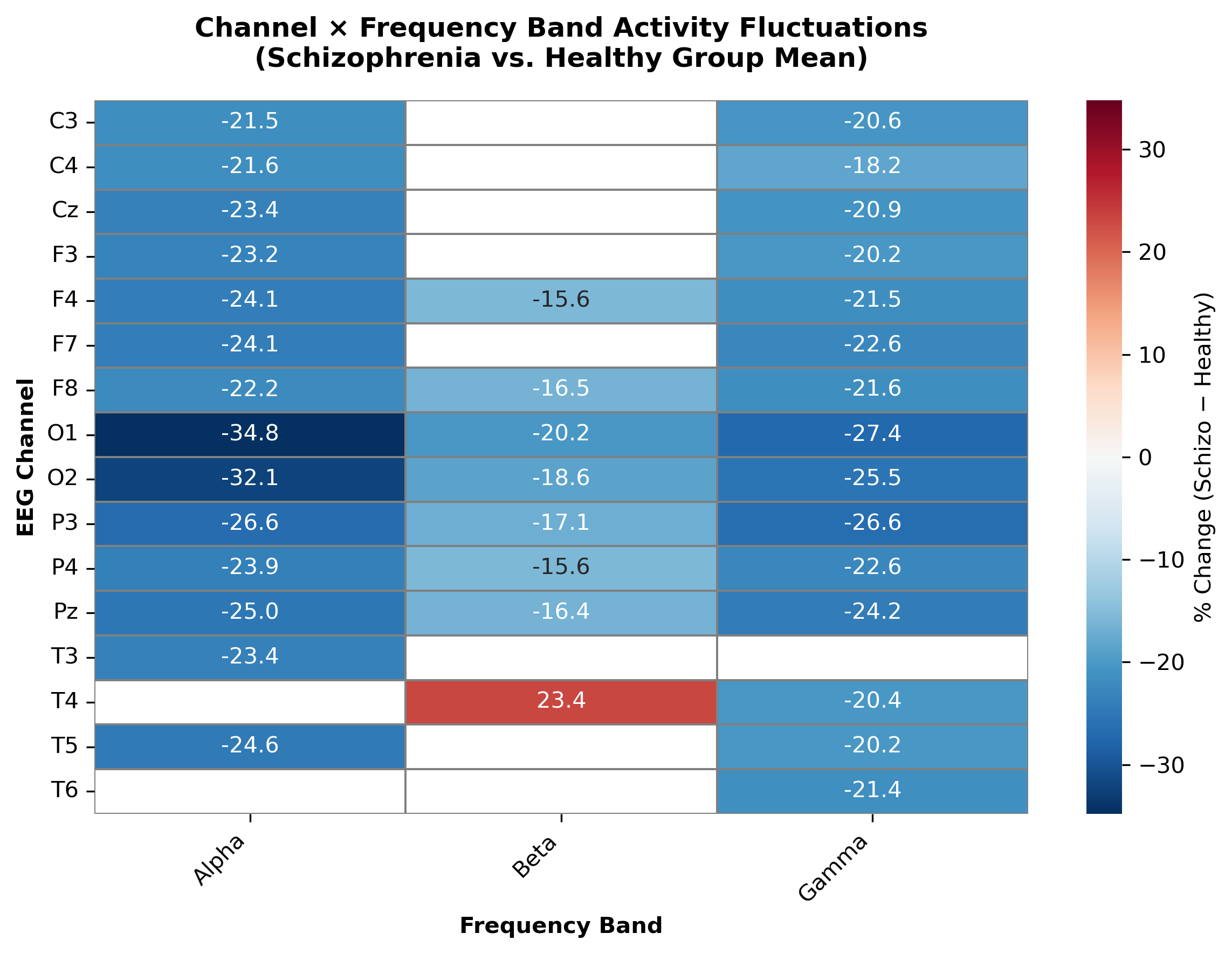}
        \centerline{\footnotesize (c) Group activity directional change (\%).}
        \label{fig:5_c}
    }
    
    \vspace{0.3cm}
    \caption{Spatial-spectral characterization and discriminative mapping of statistically significant biomarkers: (a) density distribution of discovered features mapping individual recording channels across classical EEG frequency bands, (b) localized statistical variance tracking mean $F$-scores over consolidated brain subdivisions, and (c) power spectral directional variance profiling group-level percentage activity fluctuations in schizophrenia patients compared against healthy baselines.}
    \label{fig:spatial_spectral_mapping_fig5}
\end{figure*}

The 1,255 FDR-significant WST biomarkers were heavily dominated by second-order ($S_2$; 78.5\%) over first-order ($S_1$; 21.5\%) coefficients, with zeroth-order ($S_0$) features entirely absent. Because $S_2$ captures cross-frequency coupling (CFC)\cite{b60}, while $S_1$ reflects standard power spectral density (PSD), this structural composition implies that the electrophysiological signature of schizophrenia resides primarily in disrupted temporal amplitude modulation rather than altered mean band-power\cite{b61}. Spectrally, significant biomarkers were concentrated in the gamma (57.9\%), alpha (27.6\%), and beta (14.4\%) bands, with no delta or theta features surviving correction. This slow-wave absence contrasts with traditional PSD findings\cite{b37}, indicating high cohort variability for slower alterations. Conversely, the prominent gamma enrichment aligns with documented oscillation deficits tied to parvalbumin-positive GABAergic interneuron dysfunction\cite{b62,b63}, isolating disrupted amplitude modulation of gamma rhythms as an electrical proxy for impaired microcircuit synchronization\cite{b64}. A discriminitive spatial distribution of biomarkers is shown in Fig.~\ref{fig:spatial_spectral_mapping_fig5}

\subsection{Spatial Topography and Regional Deficits}
Scalp projection of WST statistics revealed a widespread reduction in scattering energy in the schizophrenia cohort, confirming global cortical hypoactivity characterized by a distinct posterior-predominant gradient. Deficits were severe across occipital and parietal cortices but left the temporal region relatively spared. This posterior predominance aligns with resting-state literature detailing occipital alpha reductions and impaired fronto-parietal connectivity \cite{b65,b66,b67}. Conversely, temporal sparing indicates that static resting-state WST energy is less sensitive to the non-stationary, event-related dynamics typical of active hallucinations or language anomalies, which generally require task-evoked states to manifest\cite{b68,b69}. Notably, this regional profile showed high spatial concordance with the biomarker density map, positioning electrode P3—the left parietal hub of the fronto-parietal network—as the single most disrupted node generating the majority of top-ranked features.

\begin{table*}[t!]
\centering
\caption{Regional analysis of mean percentage activity change across electrode channels.}
\label{tab:regional_changes}
\begin{tabular}{llcc}
\toprule
Region & Channels & $n$ & Mean \% Change \\
\midrule
Occipital & O1, O2 & 2 & $-16.05\%$ \\
Parietal & P3, Pz, P4 & 3 & $-13.22\%$ \\
Frontal & F7, F3, F4, F8 & 4 & $-9.27\%$ \\
Central & C3, Cz, C4 & 3 & $-9.11\%$ \\
Temporal & T3, T4, T5, T6 & 4 & $-2.65\%$ \\
\midrule
Overall & All 16 & 16 & $-9.17\%$ \\
\bottomrule
\end{tabular}
\end{table*}

\begin{figure*}[t!]
    \centering
    \parbox{0.95\textwidth}{
        \centering
        \includegraphics[width=0.80\textwidth]{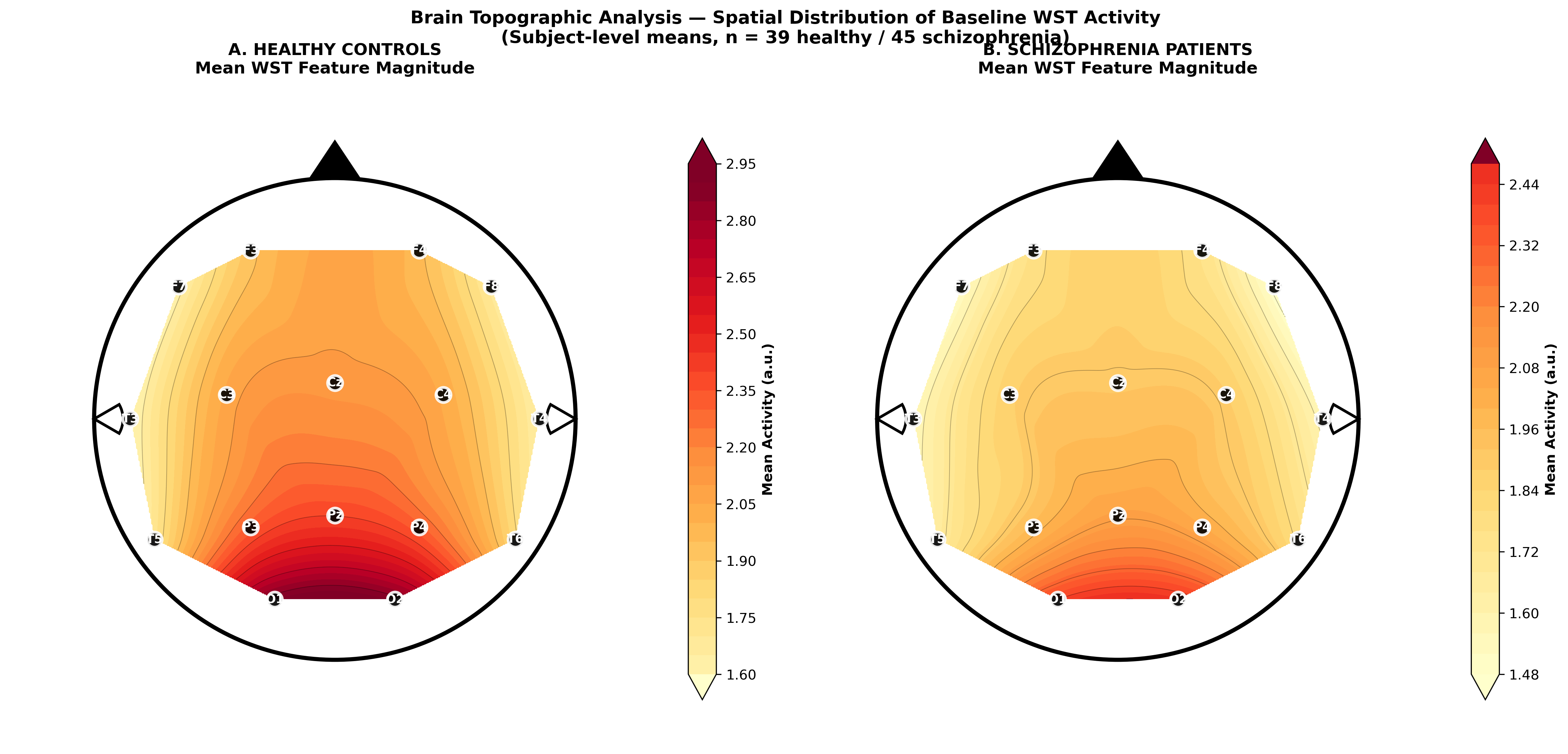}
        \vspace{0.1cm}
        \centerline{\footnotesize (a) Absolute cohort feature magnitude distributions.}
        \label{fig:6_a}
    }
    
    \vspace{0.5cm} 
    
    \parbox{0.95\textwidth}{
        \centering
        \includegraphics[width=0.8\textwidth]{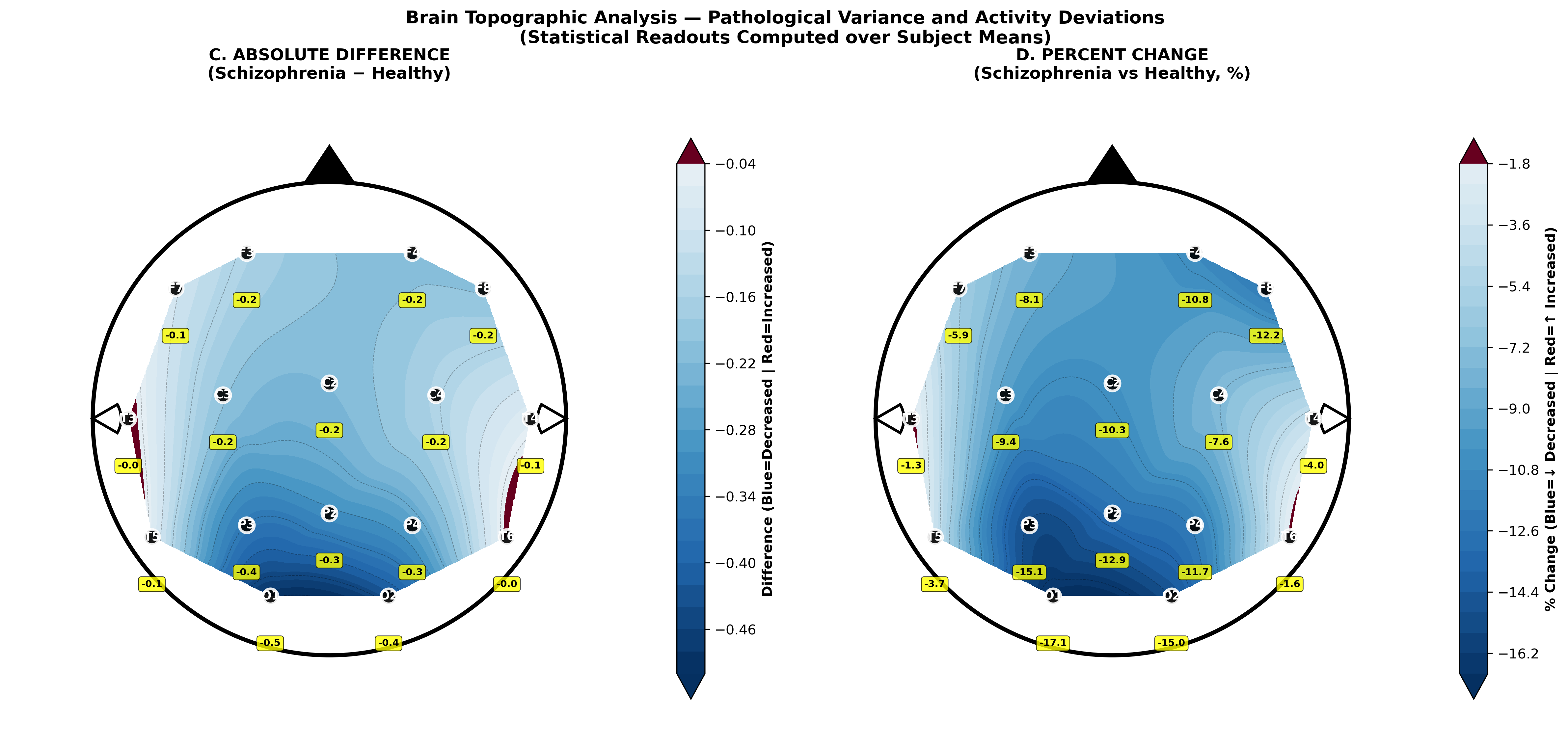}
        \vspace{0.1cm}
        \centerline{\footnotesize (b) Statistical deviations and group percentage variance.}
        \label{fig:6_b}
    }
    
    \vspace{0.4cm}
    \caption{Two-dimensional topographic mapping of spatial brain activity alterations: (a) raw baseline comparison displaying mean WST feature values across the electrode grid for Healthy Controls (left) and Schizophrenia Patients (right), and (b) diagnostic variance mapping profiling localized variations through absolute signal divergence (left) and group-level directional percentage scaling (right) across distinct cortical boundaries.}
    \label{fig:topographic_mapping_fig6}
\end{figure*}

\subsection{Classification}

Table~\ref{tab:classification_metrics} presents the complete subject-level classification performance of both classifiers under Leave-One-Subject-Out cross-validation ($n = 84$ folds):

\begin{table*}[htbp] 
\centering
\caption{Classification performance comparison between Random Forest and SVM (RBF) models.}
\label{tab:classification_metrics}
\begin{tabular}{lcc}
\toprule
Metric & Random Forest & SVM (RBF) \\
\midrule
Accuracy & $0.9048\ (0.8333 \text{ -- } 0.9643)$ & $0.7976\ (0.7024 \text{ -- } 0.8810)$ \\
Balanced Accuracy & $0.9009\ (0.8310 \text{ -- } 0.9592)$ & $0.7957\ (0.7014 \text{ -- } 0.8809)$ \\
Sensitivity & $0.9556\ (0.8837 \text{ -- } 1.0000)$ & $0.8222\ (0.7058 \text{ -- } 0.9286)$ \\
Specificity & $0.8462\ (0.7250 \text{ -- } 0.9487)$ & $0.7692\ (0.6250 \text{ -- } 0.8919)$ \\
F1-score & $0.9149\ (0.8453 \text{ -- } 0.9663)$ & $0.8132\ (0.7191 \text{ -- } 0.8932)$ \\
AUC-ROC & $0.9339\ (0.8661 \text{ -- } 0.9852)$ & $0.8860\ (0.8083 \text{ -- } 0.9501)$ \\
Cohen's $\kappa$ & $0.8072\ (0.6657 \text{ -- } 0.9248)$ & $0.5925\ (0.4014 \text{ -- } 0.7608)$ \\
MCC & $0.8110\ (0.6745 \text{ -- } 0.9258)$ & $0.5926\ (0.4044 \text{ -- } 0.7618)$ \\
\bottomrule
\end{tabular}
\end{table*}

\begin{figure*}[t][H]
    \centering
    \includegraphics[width=1.5\textwidth, height=7.5cm, keepaspectratio=true]{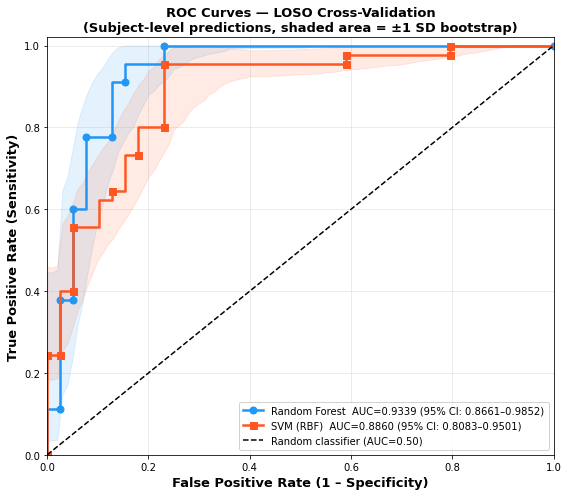}
    \caption{ROC curves for subject-independent schizophrenia classification under LOSO cross-validation}
    \label{fig:Figure__7}
\end{figure*}

\begin{figure*}[htbp]
    \centering
    
    \begin{minipage}{0.48\textwidth}
        \centering
        \includegraphics[width=0.95\linewidth]{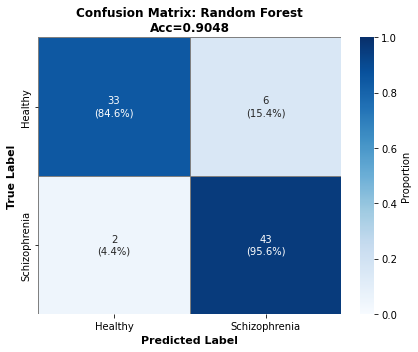}
        \\ \vspace{0.15cm}
        {\footnotesize (a) Random Forest subject-level predictions.}
        \label{fig:cm_rf}
    \end{minipage}
    \hfill
    \begin{minipage}{0.48\textwidth}
        \centering
        \includegraphics[width=0.95\linewidth]{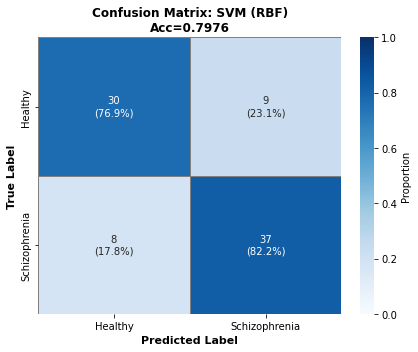}
        \\ \vspace{0.15cm}
        {\footnotesize (b) Support Vector Machine (SVM) subject-level predictions.}
        \label{fig:cm_svm}
    \end{minipage}
    
    \vspace{0.3cm}
    \caption{Confusion matrices evaluated over subject-level Leave-One-Subject-Out Cross-Validation (LOSO-CV, $n=84$).}
    \label{fig:classification_confusion_matrices}
\end{figure*}

\begin{figure*}[htbp]
    \centering
    
    \begin{minipage}{0.49\textwidth}
        \centering
        \includegraphics[width=\linewidth, height=5.5cm, keepaspectratio=false]{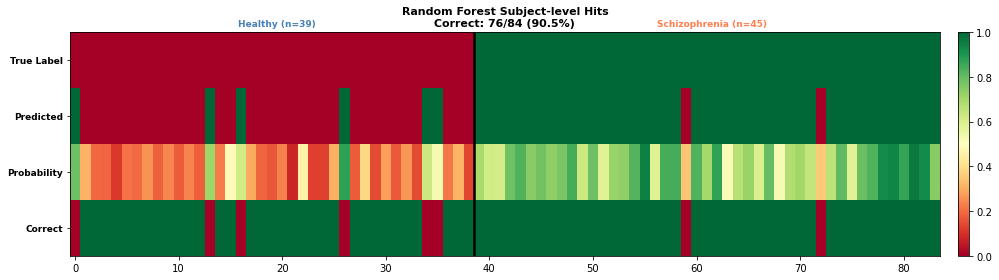}
        \\ \vspace{0.15cm}
        {\footnotesize (a) Random Forest classification profile .}
        \label{fig:profile_rf}
    \end{minipage}
    \hfill
    \begin{minipage}{0.49\textwidth}
        \centering
        \includegraphics[width=\linewidth, height=5.5cm, keepaspectratio=false]{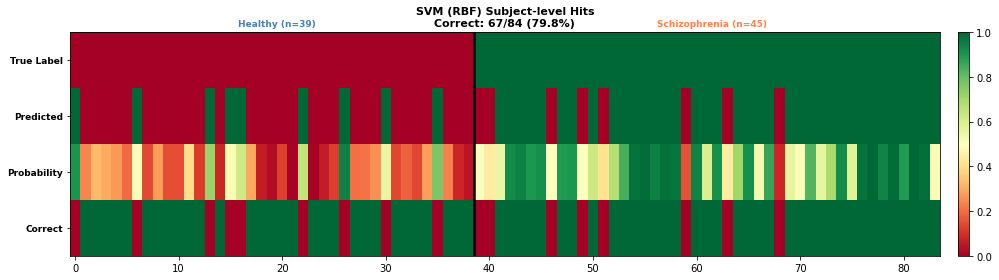}
        \\ \vspace{0.15cm}
        {\footnotesize (b) SVM (RBF) classification profile.}
        \label{fig:profile_svm}
    \end{minipage}
    
    \vspace{0.35cm}
    \caption{Per-subject classification profiles and prediction probability maps under LOSO-CV, comparison between (a) the Random Forest classifier and (b) the Support Vector Machine (SVM) with an RBF kernel.}
    \label{fig:per_subject_classification_profiles}
\end{figure*}

\begin{table*}[t!] 
\centering
\caption{Comparison of the proposed method with state-of-the-art techniques for EEG-based classification.}
\label{tab:eeg_classification_comparison}
\fontsize{7.5pt}{9.5pt}\selectfont 
\setlength{\tabcolsep}{1.8pt}       
\renewcommand{\arraystretch}{1.3} 
\begin{tabular*}{\textwidth}{@{\extracolsep{\fill}} 
  >{\raggedright\arraybackslash}p{1.6cm} 
  >{\raggedright\arraybackslash}p{2.3cm} 
  >{\raggedright\arraybackslash}p{1.7cm} 
  >{\raggedright\arraybackslash}p{1.8cm} 
  >{\raggedright\arraybackslash}p{1.5cm} 
  c c c c @{}}
\toprule
\textbf{Study \& Year} & \textbf{Feature Extraction} & \textbf{Classifier} & \textbf{Dataset Size \& Dist.} & \textbf{Validation} & \textbf{Accuracy} & \textbf{Sen.} & \textbf{Spe.} & \textbf{AUC} \\
\midrule
Jangde \& Verma (2026)\cite{b70} & Spatial-Temporal Convolutional Attention & Deep CNN & Kaggle SZ: 45, HC: 39 & 10-Fold CV & 73.98\% & 72.10\% & 75.40\% & -- \\
Siuly et al. (2020)\cite{b71} & Statistical Features + KW Test & EBT & Kaggle SZ: 49, HC: 32 & -- & 89.59\% & -- & -- & -- \\
Devia et al. (2019)\cite{b72} & ERP Features & LDA & Clinical SZ: 11, HC: 9 & -- & 71.00\% & 81\% & 59\% & -- \\
Krishnan et al. (2020)\cite{b73} & Extraction using MEMD and entropy measures & SVM/RBF & RepOD SZ: 14, HC: 14 & -- & 93\% & -- & -- & 0.9831 \\
Shoeibi et al. (2024)\cite{b74} & 1D transformer architecture & Softmax classifier & RepOD SZ: 14, HC: 14 & 10-fold CV & 97.62\% & 94.51\% & 97.74\% & -- \\
Hwang et al. (2025)\cite{b75} & Multiscale Fuzzy Entropy + Relative Power & SVM & SZ: 65, BD: 49 & -- & 78.94\% & 81.53\% & 75.51\% & -- \\
Abrar et al. (2025)\cite{b76} & Multichannel EEG processing + CAOA, RST & Proposed DL model & Kaggle & -- & 94.9\% & 93.9\% & 96.4\% & -- \\
Sravanthi et al. (2026)\cite{b77} & Variational Mode Decomposition + Multi-domain features & 9 ML and 7 optimized ML (OML) classifiers & MHRC (45 SZ / 39 HC) \& RepOD & Subject-wise LOOCV & \shortstack{96.7\% (MHRC)\\99.0\% (RepOD)} & -- & -- & -- \\
\midrule
\textbf{Proposed Method (2026)} & \textbf{Wavelet Scattering Transform (WST)} & \textbf{Random Forest (RF)} & \textbf{Kaggle SZ: 45, HC: 39} & \textbf{Subject-Level Holdout} & \textbf{90.48\%} & \textbf{95.56\%} & \textbf{84.62\%} & \textbf{0.9339} \\
\bottomrule
\end{tabular*}
\end{table*} 

\begin{figure*}[t]
    \centering
    \includegraphics[width=5\textwidth, height=7.5cm, keepaspectratio=true]{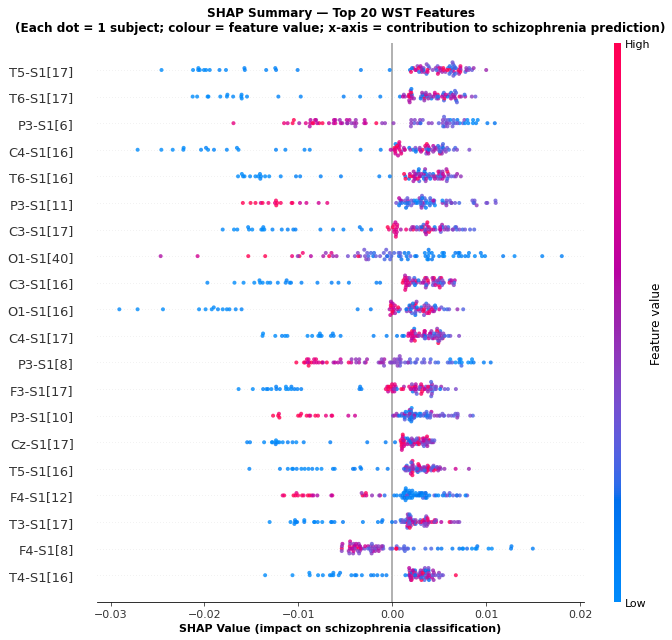}
    \caption{SHAP summary beeswarm plot of the top 20 most discriminative Wavelet Scattering Transform features}
    \label{fig:Figure__10}
\end{figure*}

\

\begin{figure*}[htbp]
    \centering
    
    \begin{minipage}{0.49\textwidth}
        \centering
        \includegraphics[width=\linewidth]{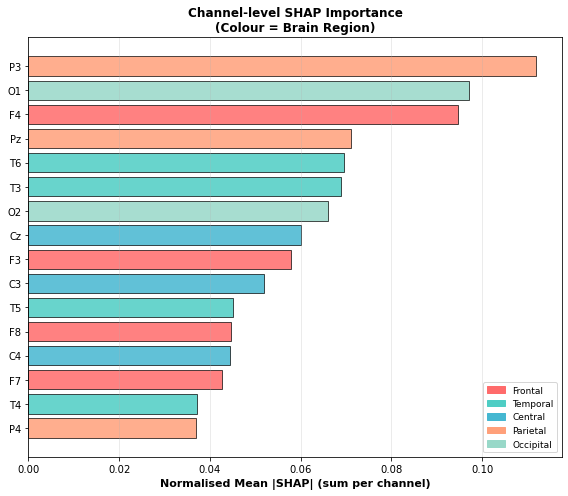}
        \\ \vspace{0.15cm}
        {\footnotesize (a) Normalized mean $|$SHAP$|$ values aggregated per channel.}
        \label{fig:channel_shap_bar}
    \end{minipage}
    \hfill
    \begin{minipage}{0.49\textwidth}
        \centering
        \includegraphics[width=0.98\linewidth]{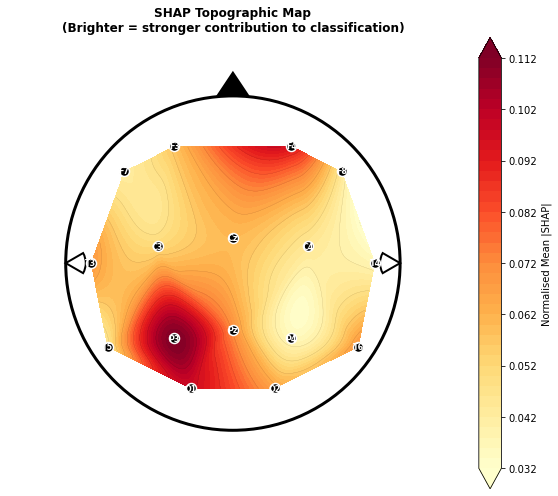}
        \\ \vspace{0.15cm}
        {\footnotesize (b) Spatial distribution of SHAP importance across the scalp.}
        \label{fig:shap_topomap}
    \end{minipage}
    
    \vspace{0.35cm}
    \caption{Spatial aggregation and channel-level SHAP importance mapping (a) Horizontal bar chart detailing feature importance summed across all scattering paths per electrode, where colors signify distinct functional brain regions. 
    (b) Interpolated SHAP topographic map visualizing global feature attribution density.}
    \label{fig:spatial_shap_analysis}
\end{figure*}

Projection of per-electrode WST statistics onto the scalp revealed a distinct, spatially organized pattern of schizophrenia-related EEG disruption, which is depicted Fig.~\ref{fig:topographic_mapping_fig6} Subject-level analysis across all electrodes demonstrated a widespread reduction in mean scattering energy in the schizophrenia group compared to healthy controls depicted in Table~\ref{tab:regional_changes}, confirming global cortical hypoactivity. Regional analysis unveiled a clear posterior-predominant gradient of disruption, with the occipital and parietal cortices showing the most severe deficits, while the temporal region was relatively spared. This posterior predominance aligns closely with established resting-state literature detailing severe occipital alpha rhythm reductions and impaired fronto-parietal connectivity during visual processing and spatial attention in schizophrenia~\cite{b61,b62,b63}. Furthermore, the relative sparing of the temporal lobe suggests that the static resting-state scattering energy captured by WST is less sensitive to the non-stationary, event-related microcircuit dynamics typical of localized auditory hallucinations and language processing abnormalities, which often manifest primarily during active symptom-related or task-evoked states rather than steady rest conditions~\cite{b64,b65}. Notably, this regional change exhibited strong spatial concordance with the biomarker density map; electrode $P3$, positioned directly over the left parietal core of the fronto-parietal network, emerged as the single most affected node, yielding the majority of the top-ranked features.

Fig. ~\ref{fig:Figure__7}, ~\ref{fig:classification_confusion_matrices}, and ~\ref{fig:per_subject_classification_profiles} comprehensively evaluate the subject-independent diagnostic performance under LOSO cross-validation, detailing the comparative ROC curves, cross-tabulated confusion matrices, and individual per-subject prediction probability profiles for both classifiers.

While existing deep-learning architectures achieve high closed-loop metrics(Table~\ref{tab:eeg_classification_comparison}) at the expense of model transparency and cross-dataset stability, our proposed WST-RF framework optimizes clinical utility by securing an exceptional subject-level sensitivity (95.56\%) and a robust AUC-ROC (0.9339). By leveraging mathematically stable $S_1$ and $S_2$ coefficients, this pipeline bypasses uninterpretable "black-box" limitations, utilizing SHAP to map decision-critical features directly onto clinically recognizable EEG nodes (P3, O1, F4) for objective diagnostic tracking.

\subsection{SHAP}
Given its superior predictive power, the full-data RF architecture was analyzed using the TreeExplainer SHAP (SHapley Additive exPlanations) framework to map feature contributions to the final classification decisions. A global SHAP summary plot of the top 20 features (Fig'~\ref{fig:Figure__10}) identified $\text{T5-S}_1[17]$, $\text{T6-S}_1[17]$, $\text{P3-S}_1[6]$, $\text{C4-S}_1[16]$, and $\text{T6-S}_1[16]$ as the five most influential metrics driving classification, heavily favoring first-order ($S_1$) scattering coefficients.

Aggregating these localized feature attributions into a channel-level importance profile (Fig.~\ref{fig:spatial_shap_analysis}-a) and an anatomical scalp topography map (Fig.~\ref{fig:spatial_shap_analysis}-b) demonstrates that the model's decision architecture is primarily prioritized around the left-parietal ($\text{P3}$), left-occipital ($\text{O1}$), and right-frontal ($\text{F4}$) regions. Finally, a Spearman rank correlation was leveraged to evaluate the spatial convergence between independent per-electrode $F$-score rankings (from the univariate statistical analysis) and the aggregated channel SHAP importance values. This revealed a moderate positive trend ($\rho_s = 0.429$, $p = 0.0969$), suggesting a localized alignment between traditional statistical group anomalies and the data-driven features prioritized by the machine learning classifier.

\section{Discussion}\label{sec12}

The dominance of second-order scattering coefficients signifies schizophrenia as a disorder of multi-scale temporal coordination rather than simple spectral power reduction and operationalizes the cross-frequency coupling (CFC) framework\cite{b61,b62}. While CFC mechanisms like theta-gamma and alpha-beta coupling are traditionally linked to active cognitive control\cite{b78}, our findings demonstrate that their baseline disruption is measurable during resting conditions, emphasizing the necessity of multi-order representations like the WST for optimal neural characterization. 

Biomarker distribution was dominated by gamma aligns with cortical deficits from parvalbumin positive GABAergic interneuron dysfunction\cite{b62,b63}. Unlike traditional PSD studies reporting elevated slow-wave power\cite{b37}, our FDR-corrected WST framework isolates robust, replicable fast-band envelope dynamics over highly variable, medication-dependent low-frequency fluctuations. Topographic analysis reveals a posterior-predominant variance maximizing at the left parietal (P3) node, indicating high WST sensitivity to microcircuit phase perturbations within parietal associative networks. This high P3 classification weight structurally aligns with localized, task-evoked phase-locked transitions documented in classic P300 wave amplitude and time-frequency analyses\cite{b79}, suggesting that baseline P3 anomalies impose an invariant computational constraint on information processing across both resting and task-driven states. Sparing frontal and temporal weights diverges from fMRI models because resting-state multi-order EEG captures tonic baseline dynamics rather than transient, episodic hallucinatory bursts\cite{b80}. This underscores neural-hemodynamic uncoupling in schizophrenia, proving that electrical microcircuit phase synchronization operates independently of macro-scale metabolic BOLD indicators\cite{b81}. 

While internally reliable, our framework requires multi-site validation across different hardware configurations and active-task paradigms to ensure real-world generalizability. Crucially, future studies must include unmedicated cohorts to isolate true pathophysiological biomarkers from antipsychotic-induced gamma-band alterations\cite{b82}. Resolving these constraints will allow longitudinal tracking of treatment efficacy and a deeper dissection of cross-order features using SHAP interaction values\cite{b83}. Ultimately, future work will extend this pipeline to determine and differentiate specific clinical subtypes of schizophrenia, moving the framework toward precision psychiatric diagnostics.

\section{Conclusion}\label{sec13}
This study establishes the Wavelet Scattering Transform (WST) as a mathematically rigorous framework that outperforms traditional spectral metrics by capturing non-linear amplitude modulations in EEG signals. The distinct dominance of second-order coefficients and gamma-band biomarkers provides clear quantitative proof that multi-scale temporal envelope dynamics, rather than static power alone, characterize resting-state schizophrenia. Spatially, these features are concentrated at the parietal P3 node, aligning with known cortical network dysfunction. Under strict Leave-One-Subject-Out (LOSO) cross-validation, a Random Forest classifier mapping this scattering space achieved a subject-independent accuracy of 90.48\% (AUC-ROC = 0.9339), verified for post-hoc transparency via SHAP attribution. Establishing a highly interpretable and reproducible signal-processing template for objective neuropsychiatric biomarker discovery, future work will extend this pipeline to determine distinct schizophrenia subtypes.

\section*{Declaration of Competing Interest}
The authors declare that they have no known competing financial interests or personal relationships that could have appeared to influence the work reported in this paper.


\section*{Data Availability Statement}
The dataset that has been used in this study is a publicly available dataset of adolescents who had been screened by psychiatrist and divided into two groups: healthy (n = 39) and with symptoms of schizophrenia (n = 45), available on Kaggle. link: \url{https://www.kaggle.com/datasets/kacharepramod/eeg-schizophrenia}.

\section*{Funding}
This research did not receive any specific grant from funding agencies in the public, commercial, or not-for-profit sectors.

.


\end{document}